\documentclass[aps,prb,preprint,showpacs,superscriptaddress]{revtex4-1}
\usepackage{amsmath}
\usepackage{graphicx}
\graphicspath{{pict/}{./}}
\usepackage{bm}%
\newcommand\pictc[5]{\begin{figure}
                       \centerline{\vspace{-1mm}
\includegraphics[width=#1\columnwidth,height=0.7\textheight,keepaspectratio]{#3}}
                       \protect\caption{\protect\label{#4} #5}\vspace{-3mm}
                    \end{figure}            }

\newcommand\pict[4][1]{\pictc{#1}{!tb}{#2}{#3}{#4}}
\newcommand\rpict[1]{\ref{#1}}

\newcounter{Fig}

\newcommand{\be}{\begin{equation}}
\newcommand{\ee}{\end{equation}}

\begin{document}
\title{Ultra-directional forward scattering by individual core-shell nanoparticles}
\author{Wei Liu}
\email{Corresponding author. Email: wli124@physics.anu.edu.au, wei.liu.pku@gmail.com}
\author{Jianfa Zhang}
\author{Bing Lei}
\author{Wenke Xie}
\author{Haotong Ma}
\author{Haojun Hu}
\affiliation{College of Optoelectronic Science and Engineering, National University of Defence
Technology, Changsha, Hunan 410073, China}


\begin{abstract}

We study the angular scattering properties of individual core-shell nanoparticles that support simultaneously both electric and optically-induced magnetic resonances of different orders.  In contrast to the approach to suppress the backward scattering and enhance the forward scattering relying on overlapping electric and magnetic dipoles, we reveal that the directionality of the forward scattering can be further improved through  the interferences of higher order electric and magnetic modes. Since the major contributing electric and magnetic responses can be tuned to close magnitudes, ultra-directional forward scattering can be achieved by single nanoparticles without compromising the feature of backward scattering suppression, which may offer new opportunities for nanoantennas, photovoltaic devices, bio-sensing and many other interdisciplinary researches.
\end{abstract}
\maketitle

\section{Introduction}

Investigations into the scattering properties of subwavelength particles are fundamentally important for researches in many directions, including sensing, optical communications, lasers, and many other interdisciplinary fields including medical and biological researches~\cite{Kerker1969_book,Bohren1983_book,Huschka2011_JACS}. Among different research fields related to the physics of particle scattering, efficient and flexible shaping of the scattering pattern is probably one of the most attractive topics, and plays a critical role for various applications including nanoantennas~\cite{Novotny2011_NP, Curto2010_Science}, photovoltaic devices~\cite{Atwater2010_NM} and so on.

As most materials have only dominant electric responses, conventional approaches to shape effectively the scattering patterns are mostly based on the engineering of the electric responses of various structures~\cite{Kerker1969_book,Bohren1983_book,Liu2014_CPB,Lukyanchuk2011_NM}. Since only electric responses are employed and manipulated, those approaches have some intrinsic limitations (see Refs.~\cite{Liu2014_CPB,Liu2012_ACSNANO} and the references therein). Although it has been well known for a long time that the scattering patterns can be more efficiently and flexibly shaped through employing the interferences of electric and magnetic modes~\cite{Kerker1969_book,Bohren1983_book,Alu2010_JN}, with outstanding examples such as Kerker's proposal~\cite{Kerker1983_JOSA} and the concept of Huygens source~\cite{Love1976_RS}, it has not attracted much attention since direct magnetic responses are rarely supported by natural materials. Recently stimulated by the vibrant field of metamaterials, a lot of structures have been demonstrated to support optically-induced magnetic resonances, which then have been employed to interfere with the electric resonances to demonstrate various kinds of scattering shaping and control~\cite{Liu2014_CPB,Jin2010_IEEE,Nieto2010_OE,Gomez-Medina2011_JN,Krasnok2011_Jetp,Liu2012_ACSNANO,Geffrin2012_NC,Filonov2012_APL,Krasnok2012_OE,Rolly2012_OE,Liu2012_PRB081407,Miroshnichenko2012_NL6459,
Fu2013_NC,Person2013_NL,Liu2013_OL2621,Staude2013_acsnano,Rolly2013_arxiv,Krasnok2014_APL,Krasnok204_nanoscale,Hancu2013_NL,Vercruysse2013_NL,Poutrina2013_OE}.

Among all the demonstrations of efficient scattering shaping involving optically-induced magnetic responses, the directional forward scattering with suppressed backward scattering obtained through overlapping electric and magnetic dipoles of the same magnitude is of fundamental importance~\cite{Liu2014_CPB,Liu2012_ACSNANO,Geffrin2012_NC,Fu2013_NC,Person2013_NL,Liu2013_OL2621}. This is due to the fact the lowest order dipolar modes are usually easiest to excite, and moreover those modes normally scatter dominantly and thus to some extent can decide the shape of the far-field scattering pattern. Also at the same time, this configuration of overlapping electric and magnetic dipoles exactly corresponds to the original proposal of Kerker~\cite{Kerker1983_JOSA} and the concept of Huygens source~\cite{Love1976_RS}.  For some applications, such as nanoantennas~\cite{Novotny2011_NP,Curto2010_Science} and photovoltaic devices~\cite{Atwater2010_NM}, ultra-directional forward scattering is required, which nevertheless cannot be achieved by the interferences of solely dipolar responses. Although it has been demonstrated recently that the directionality can be enhanced through arranging the magneto-dielectric particles in arrays~\cite{Liu2012_ACSNANO,Balanis2012_Book}, this approach sacrifices the compactness and consequently can not be used for applications such as on chip signal processing that require scattering items of small footprints. So basically ultra-directional scattering with suppressed backward scattering by individual nanoparticles would be highly preferred. It has recently been shown~\cite{Rolly2013_arxiv,Krasnok2014_APL,Krasnok204_nanoscale} that through exciting efficiently the higher order electric and magnetic modes of dielectric particles, the scattering directionality can be significantly enhanced. Unfortunately both demonstrations to some extent rely on specific incident dipole sources~\cite{Rolly2013_arxiv,Krasnok2014_APL,Krasnok204_nanoscale}. Moreover in both configurations~\cite{Rolly2013_arxiv,Krasnok2014_APL,Krasnok204_nanoscale}, it is not directly clear how to excite the modes selectively and how to control their relative phase and magnitude, which nevertheless is vitally important for further comprehensive and systematic studies.

In this paper we investigate the angular scattering properties of individual core-shell nanoparticles with incident plane waves. The core-shell nanoparticles can be tuned to support simultaneously both electric and optically-induced magnetic modes of different orders. Based on the mechanism of obtaining forward directional scattering and suppressed backward scattering by overlapping electric and magnetic dipoles of the same magnitude, it is demonstrated that the directionality of the forward scattering can be further improved when higher order electric and magnetic modes are efficiently excited. Since the major contributing electric and magnetic responses can be tuned to almost the same magnitudes, with single core-shell nanoparticles we achieve simultaneously ultra-directional forward scattering and suppressed backward scattering, which may shed new light to fields of nanoantennas, solar cells, bio-sensing and so on.

\section{Theoretical model with ideal assumptions}

Without losing the generality, we confine our discussions to incident plane waves scattered by spherical nanoparticles, which could be made single-layered or multi-layered [see Fig.~\rpict{fig1}(a)]. The plane wave is polarized along $x$ direction and propagating along $z$ direction. The scattering can be solved analytically using Mie theory~\cite{Bohren1983_book,Pena2009_CPC} and the scattering efficiency (scattering cross section divided by the cross section of the particle) is~\cite{Bohren1983_book}:
\begin{equation}
\label{efficiency}
{Q_{\rm sca}} = {2 \over {{k^2}{R^2}}}\sum\limits_{n = 1}^\infty  {(2n + 1)({{\left| {{a_n}} \right|}^2}}  + {\left| {{b_n}} \right|^2}),
\end{equation}
where R is the radius of the outmost layer of the spherical particle; $k$ is the angular wave number in the background (it is vacuum in this study); $a_n$ and $b_n$ are Mie scattering coefficients, which corresponds to \textit{n-th} order electric and magnetic moments respectively~\cite{Bohren1983_book,Wheeler2006_PRB}. At the same time, the  far-field scattering intensity (SI) is~\cite{Bohren1983_book}:
\begin{equation}
\label{SI}
{\rm SI}(\theta ,\varphi ) = {1 \over {{k^2}{l^2}}}\left[ {{{\left| {{T_1}\left( {\cos \theta } \right)} \right|}^2}{{\sin }^2}\varphi  + {{\left| {{T_2}\left( {\cos \theta } \right)} \right|}^2}{{\cos }^2}\varphi } \right],
\end{equation}
where $\theta$ and $\varphi$ are the polar angle and azimuthal angle respectively, as shown in Fig.~\rpict{fig1}(a), and $l$ is the distance between the observation point of SI and the center of the particle. The expressions for $T_{1,2}\left( {\cos \theta } \right)$ are~\cite{Bohren1983_book}:
\begin{eqnarray}
\label{T12}
\begin{array}{l}
{T_1}\left( {\cos \theta } \right) = \sum\limits_{n = 1}^\infty  {{{2n + 1} \over {n\left( {n + 1} \right)}}\left[ {{a_n}{\pi _n}\left( {\cos \theta } \right) + {b_n}{\tau _n}\left( {\cos \theta } \right)} \right]}\\
{T_2} \left( {\cos \theta } \right)= \sum\limits_{n = 1}^\infty  {{{2n + 1} \over {n\left( {n + 1} \right)}}\left[ {{a_n}{\tau _n}\left( {\cos \theta } \right) + {b_n}{\pi _n}\left( {\cos \theta } \right)} \right]}
\end{array}
\end{eqnarray}
Here ${\pi _n}\left( {\cos \theta } \right) = {{P_n^1\left( {\cos \theta } \right)} \mathord{\left/{\vphantom {{P_n^1\left( {\cos \theta } \right)} {\sin }}} \right.\kern-\nulldelimiterspace} {\sin }}\theta $,
${\tau _n}\left( {\cos \theta } \right) = {{dP_n^1\left( {\cos \theta } \right)} \mathord{\left/{\vphantom {{dP_n^1\left( {\cos \theta } \right)} d}} \right.\kern-\nulldelimiterspace} d}\theta $ and ${P_n^1\left( {\cos \theta } \right)}$ is the associated Legendre function of the first kind~\cite{Bohren1983_book}. It is easy to show that
\begin{eqnarray}
\label{parity}
\begin{array}{l}
{\pi _n}\left( { - \cos \theta } \right) = {\left( { - 1} \right)^{n + 1}}{\pi _n}\left( {\cos \theta } \right)\\
{\tau _n}\left( { - \cos \theta } \right) = {\left( { - 1} \right)^n}{\tau _n}\left( {\cos \theta } \right)
\end{array}
\end{eqnarray}
which indicates that ${\pi _n}$ and ${\tau _n}$,  ${\pi _n}$ and ${\pi _{n+1}}$, ${\tau _n}$ and ${\tau _{n+1}}$ have opposite parities with respect to $\cos \theta$.  And in the forward direction $\theta=0$:
\begin{equation}
\label{forward}
{\pi _n}\left( 1 \right) = {\tau _n}\left( 1 \right) = n(n + 1)/2.
\end{equation}
According to Eqs.~(\ref{parity})-(\ref{forward}), it is obvious that in the backward direction ($\theta=180^{\circ}$):
\begin{equation}
\label{backward}
{\pi _n}\left( { - 1} \right) + {\tau _n}\left( { - 1} \right) = 0.
\end{equation}
As we will show below, those mathematical features play a key role for scattering directionality enhancement and backward scattering suppression.

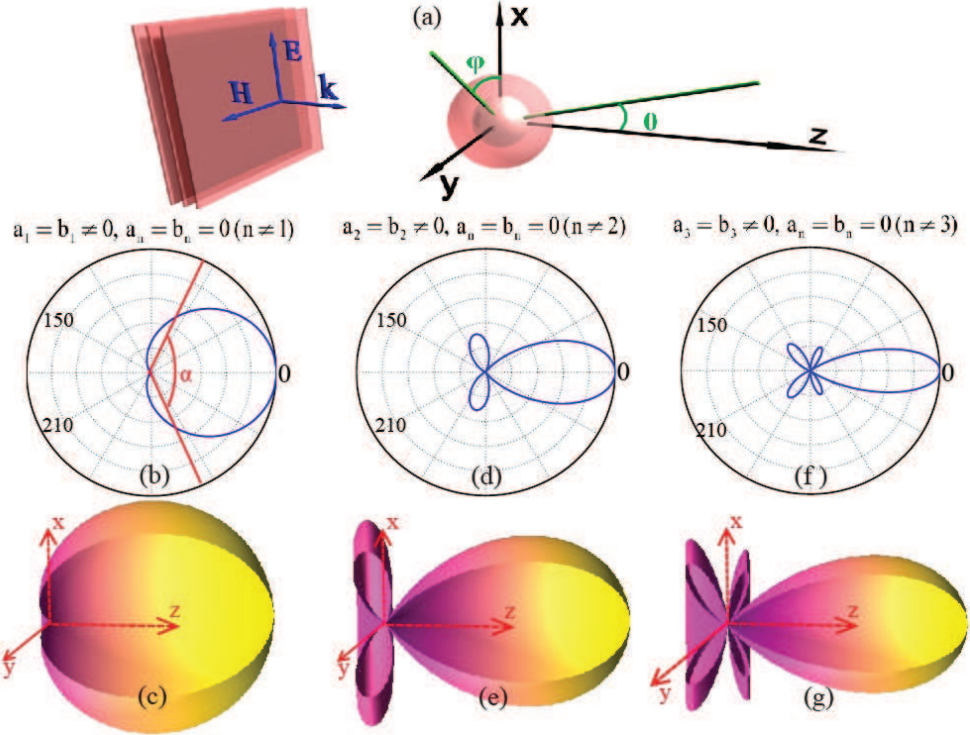
\pict[0.8]{figure1}{fig1}{\small (Color online)  (a) Scattering of an incident plane wave by a  spherical particle. The plane wave is polarized (electric field) along $x$ direction and is propagating along $z$ direction. (b)-(g) show the 2D (on a scattering plane of arbitrary fixed azimuthal angle) and 3D scattering patterns (with a part cut off for better visibility, as is the case throughout the paper) of overlapping electric and magnetic dipoles [(b) and (c)], quadrupoles [(d) and (e)] and hexapoles [(f) and (g)]. The overlapping modes are of the same order and magnitude, without exciting modes of other orders. The main lobe angular beamwidth $\alpha$ is defined as the FWHM of the SI and shown in (a).}

\subsection{Overlapping electric and magnetic dipoles}

According to Eqs.~(\ref{SI})-(\ref{T12}),  to obtain azimuthally symmetric (independent of $\varphi$) scattering patterns, it is required that $a_n=b_n$. This means that all the electric and magnetic moments of the same order should  have the same amplitudes. If this condition is satisfied, according to Eqs.~(\ref{T12})-(\ref{backward}), the SI is totally suppressed [${\rm SI}(\pi,\varphi )=0$] at the backward direction ($\theta=180^{\circ}$), and enhanced at the forward direction ($\theta=0$). Thus we can conclude that for the configuration of plane waves scattered by spherical particles, if the scattering pattern is azimuthally symmetric, the SI at the backward direction should be suppressed. The simplest case of this is that a particle supporting only dipolar electric and magnetic modes of the same magnitude $a_1=b_1\neq0$ [${a_n} = {b_n} = 0~(n > 1)$],  which is exactly what Kerker proposed~\cite{Kerker1983_JOSA} or the ideal Huygens source~\cite{Love1976_RS}.  We show the corresponding two-dimensional (2D) and three-dimensional (3D) scattering patterns in Fig.~\rpict{fig1}(b) (on a scattering plane of any fixed azimuthal angle) and Fig.~\rpict{fig1}(c) respectively. We cut off part of the 3D pattern for better visibility, as is the case throughout the paper. It is clear that the SI is zero at $\theta=180^{\circ}$ and azimuthally symmetric.  To characterize intuitively the directionality of the main scattering lobe, we define  the full width at half maximum (FWHM) of the SI as the angular beamwidth $\alpha$, as shown in Fig.~\rpict{fig1}(b). For this case, the angular beamwidth is approximately $131^{\circ}$.

\subsection{Overlapping higher order electric and magnetic modes of the same order}

It is natural to extend the case of overlapping electric and magnetic dipoles of the same magnitude to higher order modes, and according to Eqs.~(\ref{T12})-(\ref{backward}), the main features should be preserved.  Fig.~\rpict{fig1}(d) [Fig.~\rpict{fig1}(f)] and  Fig.~\rpict{fig1}(e) [Fig.~\rpict{fig1}(g)] show respectively the 2D and 3D scattering patterns of overlapping electric and magnetic quadrupoles (hexapoles) of the same magnitude, without exciting modes of other orders. It is obvious that, similar to the case of overlapping electric and magnetic dipoles [Fig.~\rpict{fig1}(b) and Fig.~\rpict{fig1}(c)], the SI is azimuthally symmetric, and totally suppressed at the backward direction.  The difference is that the angular beamwidth is significantly narrowed ($\alpha \approx 57.5^{\circ}$ and $38.8^{\circ}$ for the case of overlapping quadrupoles and hexapoles, respectively), indicating ultra-directional forward scattering. However, there is a tradeoff for better directionality with overlapping higher order modes, that other side scattering lobes arise [Fig.~\rpict{fig1}(d)-Fig.~\rpict{fig1}(g)]. This is because for overlapping dipoles [Fig.~\rpict{fig1}(b) and Fig.~\rpict{fig1}(c)], $\pi_n$ and $\tau_n$ are only perfectly in phase [the phase difference is $0$, Eq.~(\ref{forward})] at $\theta=0$ and perfectly out of phase [the phase difference is $\pi$, Eq.~(\ref{backward})] at $\theta=180^{\circ}$, without resulting in any side scattering lobes. However for overlapping higher order modes [Fig.~\rpict{fig1}(d)- Fig.~\rpict{fig1}(g)], there exist other polar angles at which $\pi_n$ and $\tau_n$ and perfectly in or out of phase~\cite{Bohren1983_book}[Eq.~(\ref{parity})], thus leading to extra side scattering lobes.

\subsection{Overlapping electric and magnetic modes of different orders}

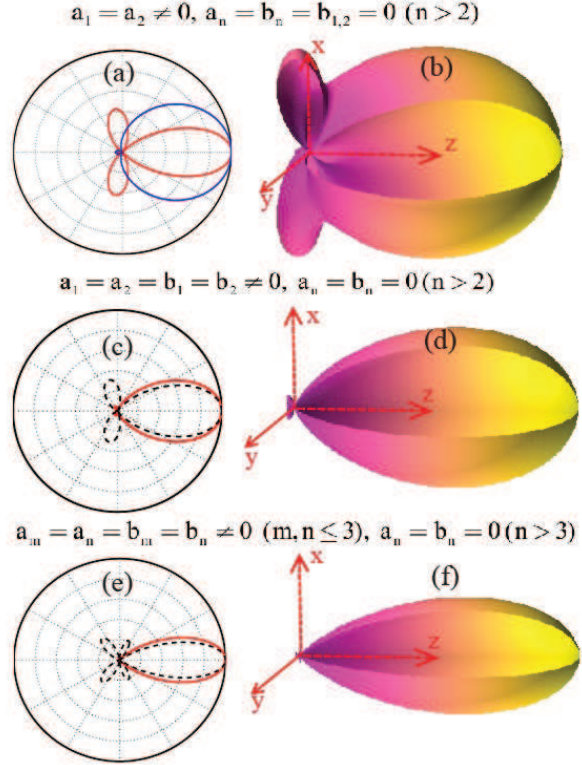
\pict[0.5]{figure2}{fig2}{\small (Color online) (a) 2D scattering pattern (red curve: scattering plane at $\varphi=0$; blue curve: scattering plane at $\varphi=\pi/2$) of overlapping electric dipole and electric quadrupole, without exciting other modes and (b) the corresponding 3D scattering pattern. (c) and (e) show respectively in red curves the 2D scattering patterns (on scattering plane of any fixed $\varphi$ as the scattering pattern is azimuthally symmetric) of overlapping electric and magnetic modes up to quadrupoles and hexapoles  of the same magnitude. The dashed black curves in (c) and (e) are the same as the curves shown in  Fig.~\rpict{fig1}(d) and Fig.~\rpict{fig1}(f) respectively. (d) and (f) show the corresponding 3D scattering patterns.}

The backward scattering suppression shown above comes from the destructive interference of the electric and magnetic modes of the same order at the backward direction~\cite{Liu2012_ACSNANO}. This originates from the mathematical features of  $\pi_n$ and $\tau_n$, which have opposite parties with respect to $\cos \theta$ [Eq.~(\ref{parity})].  According to Eq.~(\ref{parity}),  ${\pi _n}$ and ${\pi _{n+1}}$, ${\tau _n}$ and ${\tau _{n+1}}$ also have opposite parities, which means that for nanoparticles which support only electric or magnetic modes, the scattering suppression at the backward direction is also obtainable. To illustrate this principle, we show the scattering patterns of a nanoparticle which is assumed to support only electric dipole and electric quadrupole of the same magnitude in Fig.~\rpict{fig2}(a) [2D scattering pattern on the plane of $\varphi=0$ (red curve) and $\varphi=\pi/2$ (blue curve)] and in Fig.~\rpict{fig2}(b) (3D scattering pattern). It is obvious that the backward scattering has been effectively suppressed but the scattering pattern is neither azimuthally symmetric, nor has good directionality. We note here that with the assumption of $a_1=a_2$ the scattering at the backward direction is not totally suppressed [${\rm SI}(\pi,\varphi )\neq 0$]. However, if we assume that $3a_1=5a_2$, according to Eqs.~(\ref{SI})-(\ref{backward}) the scattering at the backward directional can be exactly zero. Here we show only the backward scattering suppression through overlapping electric modes of different orders and this principle can certainly apply to magnetic modes of different orders.

According to Eqs.~(\ref{SI})-(\ref{parity}) and the analysis above, it is natural to proposal that through combing the interferences between not only electric and magnetic modes of the same order, but also the modes of different orders, ultra-directional forward scattering, suppression of backward scattering and side scattering lobes can be simultaneously achieved.  To demonstrate this, we show in Fig.~\rpict{fig2}(c) in the red curve the 2D scattering pattern of overlapping dipoles and quadrupoles (both electric and magnetic) of the same magnitude (${a_1} = {a_2} = {b_1} = {b_2} \ne 0$), without involving modes of higher orders [${a_n} = {b_n} = 0~(n > 2)$]. For comparison, we show also in dashed black curve the 2D scattering pattern of overlapping quadrupoles only as already shown in Fig.~\rpict{fig1}(d). It is obvious that the side scattering lobes have been effectively suppressed without compromising much of the directionality. We also show in Fig.~\rpict{fig2}(d) the corresponding 3D scattering pattern, indicating ultra-directional forward scattering with minor side scattering lobes, which is quite different from what is shown in Fig.~\rpict{fig1}(e). Investigations have also been conducted for overlapping modes up to the third-order of hexapoles [see Fig.~\rpict{fig2}(e)-Fig.~\rpict{fig2}(f). The dashed black curve in Fig.~\rpict{fig2}(e) is the same as that shown in Fig.~\rpict{fig1}(f)] and similar conclusions can be drawn.

We should note here that in this section we study modes up to the third order of electric and magnetic hexapoles. The directionality can be further improved with suppressed backward scattering through exciting electric and magnetic modes up to higher orders.

\section{Demonstrations with realistic structures}

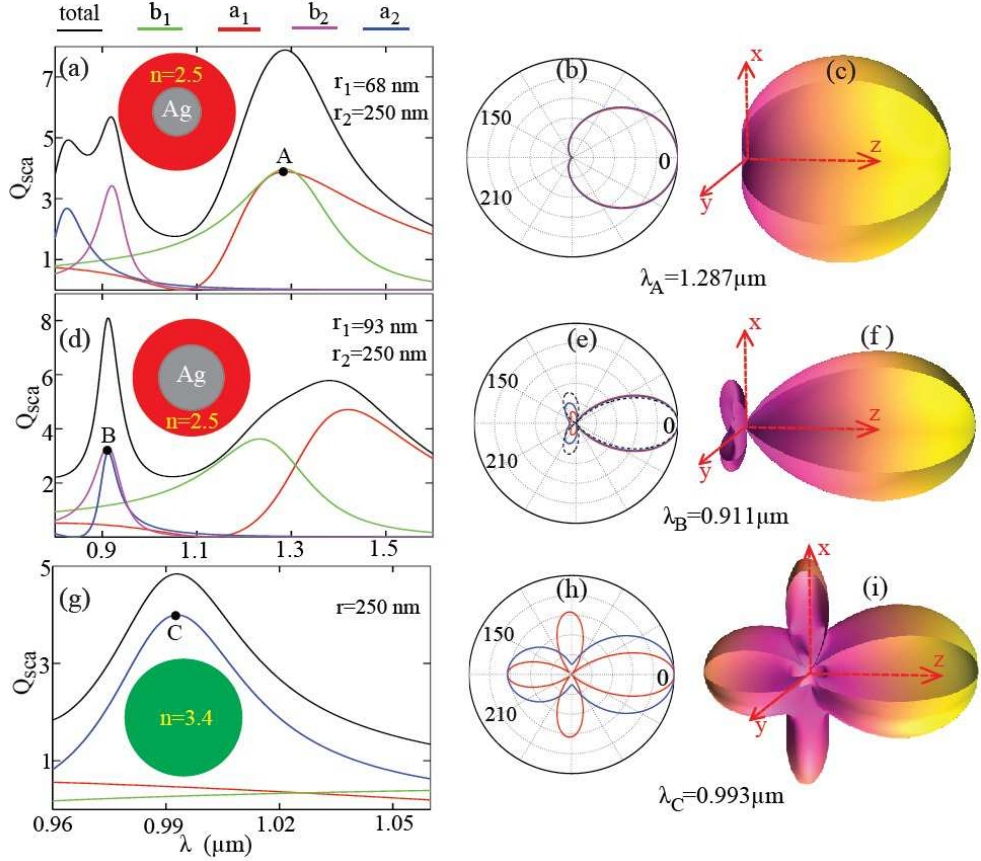
\pict[0.8]{figure3}{fig3}{\small (Color online) Scattering efficiency spectra (both total and contributions from electric and magnetic moments of different orders) of a core-shell nanoparticle with a silver core and dielectric shell of refractive index $n=2.5$  with geometric parameters of (a) $r_1=68$~nm, $r_2=250$~nm  and (d) $r_1=93$~nm, $r_2=250$~nm.  The corresponding scattering patterns are shown in (b)-(c) (at point A) and (e)-(f) (at point B) respectively. The dashed black curve in (e) are the same as the curve shown in  Fig.~\rpict{fig1}(d). (c) Scattering efficiency spectra of a homogenous dielectric sphere (refractive index $n=3.4$, radius $r=250$~nm) and the corresponding scattering patterns at point C are shown in (h) and (i). In (b), (e) and (h) red and blue curves correspond to scattering patterns on the plane of $\varphi=0$ and $\varphi=\pi/2$ respectively.}

\subsection{Overlapping electric and magnetic dipoles}

Based on our analysis and results presented in the section above, now we turn to specific structures for realistic demonstrations. Similar to what is shown in Ref.~\cite{Liu2012_ACSNANO}, we employ the core-shell spherical nanoparticle as shown as the inset of Fig.~\rpict{fig3}(a). The radii of the core and shell are $r_1$ and $r_2$ respectively. Firstly we demonstrate the simplest case of overlapping electric and magnetic dipoles.  In Fig.~\rpict{fig3}(a) we show the scattering efficiency spectra (both total and partial efficiency spectra from electric and magnetic moments of different orders) of a core-shell nanoparticle with a silver (permittivity is taken from the experimental data~\cite{Johnson1972_PRB}) core ( $r_1=68$~nm) and dielectric shell ($r_2=250$~nm) of refractive index $n=2.5$ (inset).  Apparently at point A ($\lambda=1.287$~$\mu$m) the electric dipole (characterized by $a_1$, localized surface plasmon mode of the Ag core~\cite{Liu2012_ACSNANO,Maier2007}) and the optically-induced magnetic dipole (characterized by $b_1$, cavity mode of the dielectric shell~\cite{Garixia_etxarri2011_OE,Evlykuhin2010_PRB,Kuznetsov2012_SciRep,Evlyukhin2012_NL}) overlap with the same strength. The corresponding 2D  [on the plane of $\varphi=0$ (red curve) and $\varphi=\pi/2$ (blue curve)] and 3D scattering patterns at point A are shown in Fig.~\rpict{fig3}(b) and Fig.~\rpict{fig3}(c), respectively. The scattering pattern is almost identical to that ideal case shown in Fig.~\rpict{fig1}(b)-Fig.~\rpict{fig1}(c), as according to Fig.~\rpict{fig3}(a) the scattering from other higher order modes is basically negligible.

\subsection{Overlapping electric and magnetic modes up to quadrupoles}

As a next step we increase the radius of the Ag core to $r_1=93$~nm and show the scattering efficiency spectra in Fig.~\rpict{fig3}(d). For this core-shell nanoparticle, the dipoles are now spectrally separated while the electric and magnetic quadrupoles (characterized by $a_2$ and $b_2$ respectively) overlap at point B ($\lambda=0.911$~$\mu$m) with almost the same magnitude. The corresponding scattering patterns are shown in Fig.~\rpict{fig3}(e)-Fig.~\rpict{fig3}(f).  The difference between the scattering pattern at point B [red and blues curves in Fig.~\rpict{fig3}(e)] and the scattering pattern of overlapping quadrupoles only [the dashed black curve in Fig.~\rpict{fig3}(e), as has already been shown in Fig.~\rpict{fig1}(d)] comes from the fact that at point B the scattering of dipoles [the green and red curves in Fig.~\rpict{fig3}(d)] are not negligible, which renders the scattering pattern not exactly azimuthally symmetric [red and blue curves in Fig.~\rpict{fig3}(e) do not exactly overlap]. However similar to the results shown in Fig.~\rpict{fig2}(c)-Fig.~\rpict{fig2}(d), the presence of dipoles helps to suppress the side scattering lobes [though the suppression is not as good as that shown in Fig.~\rpict{fig2}(c),  since the magnitudes of the dipoles are relatively small compared to those of the quadrupoles, as shown in Fig.~\rpict{fig3}(d)], and at the same time maintain the good directionality of the main scattering lobe [compare the red and blue curves with the dashed black curve in Fig.~\rpict{fig3}(e)]. Those features have also been exhibited by the 3D pattern shown in Fig.~\rpict{fig3}(f).

Then in Fig.~\rpict{fig3}(g) we show the scattering efficiency spectra of a homogenous dielectric (refractive index $n=3.4$) sphere. It shows that at point C  ($\lambda=0.993$~$\mu$m) the sphere supports dominantly an electric quadrupole, and  the corresponding scattering patterns are shown in Fig.~\rpict{fig3}(h)-Fig.~\rpict{fig3}(i). The scattering pattern is different from that of an ideal quadrupole, which is symmetric in the forward and backward direction~\cite{Bohren1983_book}. However the main features are shown and the discrepancy comes from the fact that at point C the scattering of dipoles [the green and red curves in Fig.~\rpict{fig3}(g)] is not negligible.   A comparison of  Fig.~\rpict{fig3}(e)-Fig.~\rpict{fig3}(f) with  Fig.~\rpict{fig3}(h)-Fig.~\rpict{fig3}(i) proves further the advantages of overlapping electric and magnetic modes for improving scattering directionality and suppressing backward scattering.

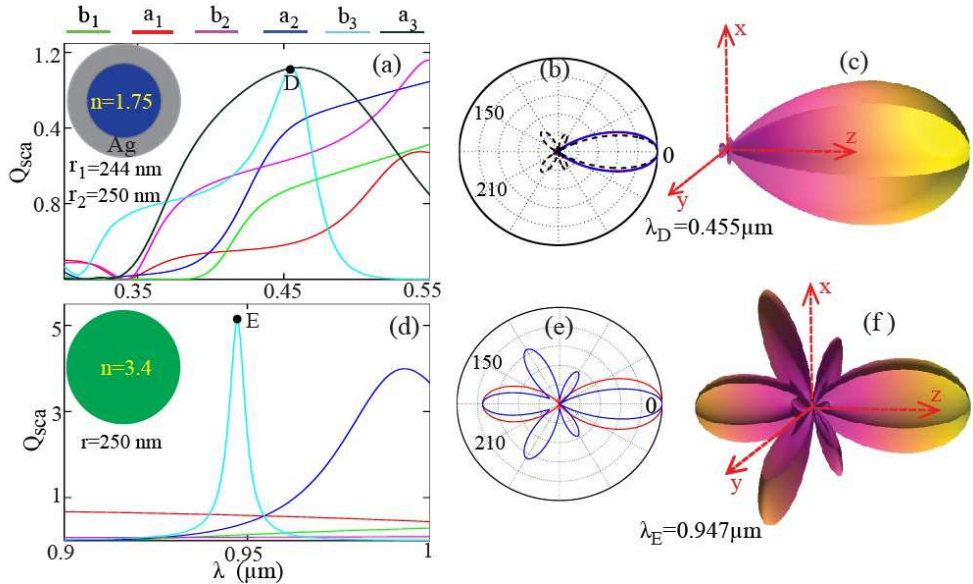
\pict[0.8]{figure4}{fig4}{\small (Color online) Scattering efficiency spectra (contributions from electric and magnetic moments up to the third order) of (a) a core-shell nanoparticle with a dielectric core ($r_1=244$~nm) of refractive index $n=1.75$ and Ag shell ($r_2=250$~nm),  and (b) a homogenous dielectric sphere (refractive index $n=3.4$, radius $r=250$~nm).  The corresponding scattering patterns are shown in (b)-(c) (at point D) and (e)-(f) (at point E) respectively. The dashed black curve in (b) is the same as the curve shown in  Fig.~\rpict{fig1}(f).  In (b) and (e) red and blue curves correspond to scattering patterns on the plane of $\varphi=0$ and $\varphi=\pi/2$ respectively.}

\subsection{Overlapping electric and magnetic modes up to hexapoles}

At the end we investigate core-shell nanoparticles that supports modes up to the third order of hexapoles. This time we employ dielectric core-Ag shell nanoparticles which can be tuned to support overlapping electric and magnetic hexapoles of the same magnitude. Fig.~\rpict{fig4}(a) shows the scattering efficiency spectra (partial efficiency spectra from electric and magnetic moments up to the third order) of a dielectric (refractive index $n=1.75$) core-Ag shell nanoparticle with $r_1=244$~nm and $r_2=250$~nm.  According to Fig.~\rpict{fig4}(a), at point D ($\lambda=0.455$~$\mu$m) the electric and magnetic hexapoles (characterized by $a_3$ and $b_3$ respectively) are tuned to overlap with  the same strength. The scattering patterns at point D are shown in Fig.~\rpict{fig4}(b)-Fig.~\rpict{fig4}(c).  For comparison, in Fig.~\rpict{fig4}(b) we also show by dashed black curve the scattering pattern of only overlapping hexapoles [the same as the curve shown in Fig.~\rpict{fig1}(f)].  In contrast to Fig.~\rpict{fig3}(e) where the side scattering lobes  have only been partly suppressed, at point D, the side scattering lobes have almost been totally eliminated [Fig.~\rpict{fig4}(b)-Fig.~\rpict{fig4}(c)],  which is quite similar to that shown in  Fig.~\rpict{fig2}(e)-Fig.~\rpict{fig2}(f). This is induced by the fact that besides the hexapoles, the quadrupoles have also been efficiently excited and are of comparable magnitudes [Fig.~\rpict{fig4}(a)]. Furthermore, at point D the good directionality has also been preserved, which leads to ultra-directional forward scattering [Fig.~\rpict{fig4}(b)-Fig.~\rpict{fig4}(c)].  Figure ~\rpict{fig4}(d) shows the scattering efficiency spectra of a dielectric sphere [the same as Fig.~\rpict{fig3}(g) but at a different spectra regime] and obviously that at point E ($\lambda=0.947$~$\mu$m) a dominant magnetic hexapole is supported. Its scattering patterns are shown in Fig.~\rpict{fig4}(e)-Fig.~\rpict{fig4}(f), which are contrastingly different from those shown in Figure ~\rpict{fig4}(b)-(c), which proves again that through combining the interferences between not only electric and magnetic modes of the same order, and but also between modes of different orders, the scattering directionality can be efficiently improved with the feature of suppressed backward scattering preserved.

\section{Conclusions}

To conclude, we study the scattering configuration of plane waves scattered by individual core-shell nanoparticles which are tuned to support simultaneously both electric and optically-induced magnetic modes of different orders.
In contrast to the forward directional scattering and suppressed backward scattering obtained through overlapping electric and magnetic dipoles with the same magnitude, we reveal that through interferencing other higher order electric and magnetic modes which have been efficiently excited, the scattering directionality can be further improved. Since the major contributing electric and magnetic responses can be tuned to close magnitudes, by single nanoparticles we can achieve simultaneously ultra-directional forward scattering, and suppression of extra scattering lobes and backward scattering.

We note here that we confine our discussions to two-layered core-shell nanoparticles, which offers us sufficiently freedom to tune the electric and magnetic modes of a specific order to overlap with the same magnitude and central resonant wavelength [as shown in Fig.~\rpict{fig3}(a), Fig.~\rpict{fig3}(d) and Fig.~\rpict{fig4}(a)].  However for two-layered nanoparticles, we cannot tune the relative strength and phase of all the modes at the same time, which imposes some constraints on more efficient and flexible scattering shaping [\textit{e.g}., as shown in Fig.~\rpict{fig3}(d), the contributions from dipoles are too small compared to those from quadrupoles and consequently the side scattering lobes cannot be totally suppressed, as shown in Fig.~\rpict{fig3}(e)].  We expect that more freedom for resonance tuning (including magnitude and phase) and thus more flexibilities for scattering shaping can be obtained by employing core-shell nanoparticles with more layers. Moreover we should keep in mind that the ultra-directional forward scattering can be further collimated through exciting higher order modes than hexapoles or arranging the core-shell nanoparticles in arrays. It is worth noticing that the mechanism we reveal in this paper is not constrained to spherical nanoparticles and can certainly be extended to particles of other shapes, which is quite promising for various applications in the fields of nanoantennas, solar cells, bio-sensing and so on.

\section{Acknowledgements}

We thank Andrey E. Miroshnichenko, Dragomir N. Neshev, Rupert F. Oulton, and Yuri S. Kivshar for valuable discussions.  We also acknowledge the financial support from the National Natural Science Foundation of China (Grant No. 11304389 and 61205141) and the Basic Research Scheme of College of Optoelectronic Science and Engineering, National University of Defence Technology.


\end{document}